\documentclass[preprint,aps,amsmath,superscriptaddress,nofootinbib]{revtex4}
\usepackage{bm}
\usepackage{epsfig}

\def\bsigma{\mbox{\boldmath $\sigma$}}

\newcommand{\nn}{\nonumber} 

\newcommand{\bea}{\begin{eqnarray}}
\newcommand{\eea}{\end{eqnarray}}

\newcommand{\mpi}{m_{\pi}}
\newcommand{\md}{m_{D}}
\newcommand{\mds}{m_{D^*}}
\newcommand{\mx}{m_{X}} 
\newcommand{\mdds}{M_{DD*}}
\newcommand{\vd}{v_D}

\newcommand{\mev}{\textrm{MeV}}

\newcommand{\bd}{ \bm{  D  } }
\newcommand{\bdbar}{ \bm{  {\bar D } } }

\begin{document}



\title{Pion Interactions in the $X(3872)$} 

\author{S. Fleming\footnote{Electronic address:     fleming@physics.arizona.edu}}
\affiliation{Department of Physics, 
        University of Arizona,
	Tucson, AZ 85721
	\vspace{0.2cm}}
	
\author{M. Kusunoki\footnote{Electronic address: masa@physics.arizona.edu}}
\affiliation{Department of Physics, 
        University of Arizona,
	Tucson, AZ 85721
	\vspace{0.2cm}}

\author{T. Mehen\footnote{Electronic address: mehen@phy.duke.edu}}
\affiliation{Department of Physics, 
	Duke University, Durham,  
	NC 27708\vspace{0.2cm}}
\affiliation{Jefferson Laboratory, 
	12000 Jefferson Ave., 
	Newport News, VA 23606\vspace{0.2cm}}
\affiliation{Center for Theoretical Physics, Massachusetts Institute of Technology, 
Cambridge, MA 02139	\vspace{0.2cm}}

\author{U. van Kolck\footnote{Electronic address: vankolck@physics.arizona.edu}}
\affiliation{Department of Physics, 
        University of Arizona,
	Tucson, AZ 85721
	\vspace{0.2cm}}
\affiliation{Kernfysisch Versneller Instituut,
        Rijksuniversiteit Groningen,
	Zernikelaan 25, 9747 AA Groningen, The Netherlands
        \vspace{0.2cm}}
\affiliation{Instituto de F\'\i sica Te\'orica,
        Universidade Estadual Paulista,
	Rua Pamplona 145, 01405-900 S\~ao Paulo, SP, Brazil
	\vspace{0.2cm}}

\date{\today\\ \vspace{1cm} }


\begin{abstract}

We consider 
pion interactions in an effective field theory of the 
narrow resonance $X(3872)$, assuming it is 
a weakly bound molecule of the charm mesons
$D^{0} \bar D^{*0}$ and $D^{*0} \bar D^{0}$. 
Since the hyperfine splitting of 
the $D^{0}$ and $D^{*0}$ is only 7
MeV greater than the neutral pion mass, pions can be produced near
threshold and are  non-relativistic. 
We show that 
pion exchange can be treated in
perturbation theory and 
calculate the next-to-leading-order correction
to the partial decay width  $\Gamma[X\to D^0 \bar D^{0} \pi^0]$.

\end{abstract}

\maketitle

\newpage

\section{Introduction}
The idea that the recently discovered 
$X(3872)$ is a shallow molecular bound state of $D^{*0} \bar{D^0}$ and 
$\bar{D}^{*0} D^0$ mesons is extremely attractive and 
has motivated numerous calculations of $X(3872)$ properties 
using effective-range theory, for a review see Ref.~\cite{Voloshin:2006wf}. 
Going beyond this approximation
requires including effects from dynamical pion exchange. 
The goal of this paper is to develop an effective theory of non-relativistic 
$D$ mesons and pions 
that can be used to compute properties of the $X(3872)$ 
systematically at low energies. Due to the accidental
nearness of the $D^*$-$D$ hyperfine splitting and the pion mass, 
pion exchanges are characterized by an anomalously small scale compared 
to what is usually the case in nuclear physics~\cite{Suzuki:2005ha}.
We argue in this paper that, unlike in conventional nuclear physics, 
these effects can be treated using perturbation theory and compute 
the decay $X \to D^0 \bar{D}^0 \pi^0$ to next-to-leading order (NLO) 
in the effective theory.

We begin by reviewing the current  experimental understanding of the 
$X(3872)$. The $X(3872)$ is a narrow resonance discovered by the Belle
collaboration \cite{Choi:2003ue} in electron-positron collisions 
through the decay 
$B^{\pm}\to XK^{\pm}$ followed by the decay $X\to J/\psi\,\pi^+\pi^-$.
Its existence has been confirmed by the CDF and D\O  ~collaborations
through its inclusive production in proton-antiproton collisions
\cite{Acosta:2003zx, Abazov:2004kp} and by the Babar collaboration
through the discovery mode $B^\pm\to XK^\pm$ \cite{Aubert:2004ns}.
The combined averaged mass of the $X(3872)$ measured by these experiments is 
\cite{Olsen:2004fp}
\begin{eqnarray}
 m_X = 3871.2 \pm 0.5  \; {\rm MeV}.
\label{mX}
\end{eqnarray}
Note that the mass of the $X(3872)$ is quite close to the $D^{0}\bar D^{*0}$
threshold at $3871.81\pm 0.36$ MeV \cite{Cawlfield:2007dw}.
The Belle collaboration has placed an upper limit on the width of
the $X(3872)$ \cite{Choi:2003ue}:
\begin{eqnarray}
 \Gamma_X  < 2.3 \; {\rm MeV} \;\; (90 \%\;{\rm C.L.}).
\label{GammaX}
\end{eqnarray}

The $X(3872)$ has also been observed in  
the decays $X\to J/\psi\,\pi^+\pi^-\pi^0$ 
and $X\to J/\psi\,\gamma$ \cite{Abe:2005ix}.
The ratio of branching fractions for the three- and two-pion final states is
~\cite{Abe:2005ix} 
\bea\label{isoval}
\frac{{\rm Br}[X\to J/\psi \,\pi^+ \pi^- \pi^0]}
     {{\rm Br}[X\to J/\psi \, \pi^+ \pi^-]}
= 1.0 \pm 0.4 \pm 0.3 \, .
\eea
Since these decays are thought to proceed through $J/\psi \,\rho$ for the 
$J/\psi \,\pi^+ \pi^-$ final state, and through $J/\psi \, \omega$ for
the $J/\psi \,\pi^+ \pi^- \pi^0$ final state, the ratio in Eq.~(\ref{isoval}) 
indicates a large violation of isospin invariance. 
A near-threshold enhancement in 
$D^0 \bar{D}^0 \pi^0$ has been observed in $B\to D^0 \bar{D}^0 \pi^0 K$ 
decays~\cite{Gokhroo:2006bt}. This is the first evidence for the decay 
$X\to D^0 \bar{D}^0 \pi^0$, though the peak of the observed resonance
in Ref.~\cite{Gokhroo:2006bt} is at 
$3875.2 \pm 0.7^{+0.3}_{-1.6}\pm 0.8 \, \mev$,
which is 2$\sigma$ above the world-averaged  $X(3872)$ mass.
The branching ratio for $X\to D^0 \bar{D}^0 \pi^0$
observed in Ref.~\cite{Gokhroo:2006bt} is $8.8^{+3.1}_{-3.6}$  larger than 
the discovery  mode $X\to J/\psi \, \pi^+ \pi^-$. 
The Babar collaboration has established 
${\rm Br}[X\to J/\psi\,\pi^+\pi^-] > \, 0.042$ 
at $90\%$ C.L.~\cite{Mohanty:2005dm,Aubert:2005vi}.
Various upper limits have been placed on the product of 
${\rm Br}[B^\pm\to X K^\pm]$ and other
branching fractions of the $X(3872)$
including $D^{0}\bar D^{0}$, $D^+D^-$~\cite{Abe:2003zv},
$\chi_{c1}\gamma$, $\chi_{c2}\gamma$, $J/\psi\,\pi^0\pi^0$ 
\cite{Abe:2004sd},
and $J/\psi\, \eta$ 
\cite{Aubert:2004fc}. 
Upper limits have also been placed on the partial widths
for the decay of $X(3872)$ into $e^+e^-$ 
\cite{Yuan:2003yz, Dobbs:2004di}
and into $\gamma\gamma$ \cite{Dobbs:2004di}.

The possible $J^{PC}$ quantum numbers of the $X(3872)$ 
have been examined. 
The observation of $X \to J/\psi\,\gamma$ 
establishes $C=+$. 
This is consistent with 
the shape of the $\pi^+\pi^-$ invariant mass distributions 
\cite{Choi:2003ue, Aubert:2004ns, Abulencia:2005zc}.
Belle's angular distribution analysis of  
$X\to J/\psi\,\pi^+\pi^-$ favors $J^{PC}=1^{++}$ 
\cite{Abe:2005iy}.
A recent CDF analysis~\cite{Abulencia:2006ma} finds that 
$J/\psi \, \pi^+ \pi^- $ angular distributions are only 
consistent with $J^{PC} = 1^{++}$ and $2^{-+}$.

The quantum numbers 
$J^{PC} = 1^{++}$ arise if  
the $X(3872)$ is a $C=+$, $S$-wave molecular bound state
of $D^0\bar{D}^{*0} + \bar{D}^0 D^{*0}$. 
The possibility of a shallow molecular state is motivated by the proximity 
of the $X(3872)$ to the $D^{0} \bar{D}^{*0}$ threshold and  naturally 
explains the large isospin violation observed in pion decays and 
the dominance of the $D^0 \bar{D}^0 \pi^0$ decay mode. 
The narrow width and the non-observation of decays
such as $X\to \chi_c \gamma$ are highly unusual for a conventional charmonium 
state above the $D \bar{D}$ threshold. 
{}From the mass in Eq.~(\ref{mX}) and the recent measurement of the $D^0$ mass
in Ref.~\cite{Cawlfield:2007dw} one infers a binding energy
\begin{eqnarray}
 E_X &=&  \md + \mds - \mx \nonumber \\
     &=&   0.6 \pm 0.6 \; \mev.
\label{eb} 
\end{eqnarray}
This favors a bound-state interpretation
of the $X(3872)$, however, because of the large uncertainty, the mass alone
cannot rule out  a resonance or ``cusp'' near the $D^0\bar{D}^{*0}$ 
threshold ~\cite{Bugg:2004rk}.
In this paper we will assume the $X(3872)$ is a molecular bound state, 
though our method can be extended to the case where the $X(3872)$ is 
a shallow resonance.
For other interpretations, see Refs.~\cite{Barnes:2003vb, Eichten:2004uh,
Tornqvist:2004qy,Wong:2003xk,Swanson:2003tb,
Close:2003mb, Li:2004st,Seth:2004zb,Maiani:2004vq,Navarra:2006nd,
Ishida:2005em,Vijande:2004vt, Bauer:2005yu,Kim:2006fa,Swanson:2006st, 
Colangelo:2007ph}.
A recent review can be found in Ref.~\cite{Zhu:2007wz}. 

The interpretation as a $D D^*$ molecule is particularly predictive
because the small binding energy implies that the molecule has universal
properties that are determined by the binding energy
\cite{Voloshin:2003nt,Voloshin:2005rt,Braaten:2003he,Braaten:2004rw,Braaten:2004fk,Braaten:2004ai}.
The small binding energy can be further exploited through factorization 
formulae for production and decay rates of the $X(3872)$
\cite{Braaten:2005jj, Braaten:2005ai}. Voloshin calculated the decays 
$X\to D^{0}\bar D^{0} \pi^0$~\cite{Voloshin:2003nt} and 
$X\to D^{0}\bar D^{0} \gamma$~\cite{Voloshin:2005rt}
using the universal wavefunction of the molecule.

\begin{figure}
\begin{center}
\includegraphics[width=5cm]{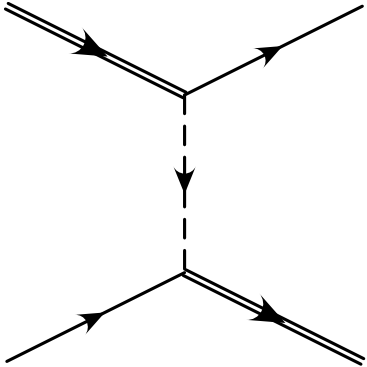}
\caption{ One-pion exchange diagram for
$D^{*0} \bar D^{0} \to D^{0} \bar D^{*0}$ scattering. 
The single and double lines represent the spin-0 
and spin-1 $D$ mesons, respectively.
The dashed line represents the  $\pi^0$. 
\label{pionex}}
\end{center}
\end{figure}

The main purpose of this paper is to consider the effect of $\pi^0$
exchange on the properties of the $X(3872)$. Consider the one-pion exchange
contribution to $D^{*0} \bar{D}^0 \to  D^0  \bar{D}^{*0}$ scattering 
depicted in Fig.~\ref{pionex}.
This leads to an amplitude 
\bea\label{ope}
\frac{g^2}{2f_\pi^2}
\frac{{\vec{\epsilon}}^{\,\, *} \cdot \vec{q} \, \vec{\epsilon}\cdot \vec{q}}
     {\vec{q}^{\,2} - \mu^2}\, ,
\eea
where $g$ is the $D$-meson axial (transition) coupling, 
$f_\pi$ is the pion decay constant,
$\vec \epsilon$ and ${\vec{\epsilon}}^{\,\, *}$ are the polarization vectors 
of the incoming and outgoing $D^*$ mesons, respectively,
and $\vec{q}$ is the momentum transfer. 
The scale $\mu$ appearing in the propagator denominator is given by 
$\mu^2 = \Delta^2 -m_\pi^2$, where $\Delta$ is the $D^*$-$D$ hyperfine 
splitting and $m_{\pi}$ is the neutral pion mass.
The hyperfine splitting, $\Delta$, appears in the pion propagator because 
the exchanged pion carries energy $q^0 \simeq \Delta$ as well as 
momentum $\vec{q}$.
Note that $\mu$ is anomalously small, $\mu \approx 45\, {\rm MeV}$, 
because of the nearness of $\Delta = 142\, {\rm MeV}$
and $m_{\pi}=135 \, {\rm MeV}$. 
This suggests that pions generate anomalously long-range effects
and should be included as explicit degrees of freedom
in the description of the molecule, if
the binding energy in Eq.~(\ref{eb}) is not much smaller 
than its upper limit.

The pion interactions in the $D$ and $D^*$ system were
quantitatively analyzed using a
one-pion-exchange potential model by Tornqvist \cite{Tornqvist:1993ng},
who actually predicted a $D\bar{D}^*$ bound state (deuson) with 
a mass close to the observed $X(3872)$.
After the discovery of the $X(3872)$, Swanson~\cite{Swanson:2003tb} considered
a potential model that includes both a one-pion-exchange
potential and a quark-exchange potential and
found a weakly bound state in the S-wave $J^{PC}=1^{++}$ channel.
These authors 
worked in the isospin limit and used isospin-averaged pion masses and 
hyperfine splittings and obtained
long-range Yukawa-like potentials~\cite{Tornqvist:1993ng}.
Note that the effective mass term in the
propagator in Eq. (\ref{ope}) has the opposite
sign from what one typically obtains from meson exchange. 
This leads to a $\pi^0$-exchange potential in position space which is
oscillatory rather than Yukawa-like, 
as pointed out by Suzuki~\cite{Suzuki:2005ha}. 

A central point of this paper is that the effect of $\pi^0$ exchange can be 
dealt with using perturbation theory. Naive dimensional analysis of the 
relative size of two-pion and one-pion exchange graphs yields the ratio
\bea\label{Xexp}
\frac{g^2 M_{DD^*} \mu}{4 \pi f_\pi^2} \approx \frac{1}{20}-\frac{1}{10} \, ,
\eea
where $M_{DD^*}$ is the reduced mass of the $D$ and $D^*$ and we have set 
$g = 0.5 -0.7$~\cite{Ahmed:2001xc, Anastassov:2001cw,Fajfer:2006hi}.
This is in contrast with two-nucleon systems where a similar estimate 
yields~\cite{Kaplan:1998tg, Kaplan:1998we}
\bea\label{NNexp}
\frac{g_A^2 M_N m_\pi}{8 \pi f_\pi^2} \approx \frac{1}{2} \, ,
\eea
where $g_A=1.25$ is the nucleon axial coupling and $M_N$ is the nucleon mass.
A perturbative treatment of pions fails in the $^3S_1$ channel where 
iteration of the spin-tensor force yields large corrections at 
next-to-next-leading order (NNLO)~\cite{Fleming:1999bs,Fleming:1999ee}. 
This is in part due to the large expansion parameter in Eq.~(\ref{NNexp}) 
and in part due to large numerical coefficients appearing in the 
NNLO calculation. The amplitude in Eq.~(\ref{ope}) also gives rise  
to a spin-tensor force and one may worry that the perturbative treatment 
of pions will fail. However, even if large NNLO coefficients like those found 
in Ref.~\cite{Fleming:1999bs,Fleming:1999ee}
appear in similar diagrams for the $X(3872)$, the expansion parameter 
in Eq.~(\ref{Xexp}) is 
small enough that one can reasonably expect perturbation theory to work.

In this paper, 
we derive an effective field theory of
the $D^{0}\bar D^{*0}$ and $D^{*0}\bar D^{0}$
interacting with neutral pions near 
the $D^{0}\bar D^{*0}$ threshold. This theory is very similar in structure 
to the KSW theory of $NN$ interactions 
in Ref.~\cite{Kaplan:1998tg, Kaplan:1998we}
where a leading-order (LO) contact interaction is summed to all orders 
in perturbation
theory to produce a bound state at LO and pion exchange is treated
perturbatively.~\footnote{A pionless effective theory of shallow nuclear bound
states in which the leading non-derivative contact interaction is resummed to 
all orders was first proposed in 
Ref.~\cite{vanKolck:1998bw}.  For a similar theory of the $X(3872)$ see 
Ref.~\cite{AlFiky:2005jd}. }
A novel feature of the effective theory for the $X(3872)$ is that the 
hyperfine splitting of the $D^{0}$ and $D^{*0}$ is only 
7 MeV above the $\pi^0$ mass and thus the pions are 
included as non-relativistic particles. In this paper we focus on the decay
$X\to D^0 \bar{D}^0 \pi^0$. Our results are easily extended to 
$X \to D^0 \bar{D}^0 \gamma$.
At LO our theory reproduces Voloshin's calculations using 
effective-range theory~\cite{Voloshin:2003nt,Voloshin:2005rt}. 
We then compute the NLO corrections to the decay width.
These include effective-range corrections as well as calculable 
non-analytic corrections from 
$\pi^0$ exchange. We find that non-analytic calculable corrections 
from pion exchange are negligible and the NLO correction is dominated 
by contact interaction contributions.

This paper is organized as follows. In Section II, we give the Lagrangian 
and discuss power counting in our theory. In Section III, we describe 
our calculation of the partial width $\Gamma[X\to D^0 \bar{D}^0 \pi^0]$. 
In Section IV, we summarize and conclude. 
Appendix A describes how our Lagrangian is derived by integrating out 
the scales $m_\pi$ and 
$\Delta$ from heavy-hadron chiral perturbation theory 
(HH$\chi$PT)~\cite{Wise:1992hn,Burdman:1992gh,Yan:1992gz}. 
Appendix B gives the results of evaluating the individual NLO diagrams 
for the decay amplitude and the wavefunction renormalization.

While this work was being completed, a related preprint~\cite{Braaten:2007ct} 
appeared
which analyzed the effects of light-meson exchange on a bound state
of heavy mesons near a three-meson threshold. This work used a scalar-meson
model and calculated the entire line shape of the resonance to second
order in the heavy-light meson coupling. Our work is complimentary to that 
of Ref.~\cite{Braaten:2007ct} in that we do not use a model but rather a 
Lagrangian 
that is directly relevant to the $X(3872)$ and we go to higher order 
in the heavy-light meson coupling, where renormalization requires 
the introduction of higher derivative contact operators.
On the other hand, we do not calculate the full line shape but work at
the resonance peak where a Breit-Wigner is a suitable approximation.

\section{Lagrangian and Power Counting}

The mass of the $X(3872)$ in Eq.~(\ref{mX}) is 
extremely close to the $D^0 \bar D^{*0}$ threshold.
Assuming that the $X(3872)$ is a hadronic
molecule whose constituents are a superposition of the 
$D^{0} \bar D^{*0}$ and $D^{*0} \bar D^{0}$,
the $X(3872)$ binding energy is given by Eq.~(\ref{eb}). 
For reasons stated earlier, our calculations assume positive binding energy
and a molecular interpretation of the $X(3872)$. 
The upper bound on the typical momentum
of the $D$ and $\bar D^*$ in the bound state is then
$\gamma \equiv ( 2 \mdds E_X )^{1/2} \le 48 \, \mev$, 
where $\mdds$ is the reduced mass 
of the $D^0$ and $\bar D^{*0}$. For this binding momentum the 
typical velocity of the $D$ and $D^*$ is approximately 
$\vd \simeq ( E_X / 2 \mdds )^{1/2} \lesssim 0.02$, 
and both the $D$ and $D^*$ are clearly non-relativistic. 
We will use non-relativistic fields for the $D$ and $D^*$.

The pion degrees of freedom are also treated non-relativistically.  
The maximum energy of the pion emitted in the decay 
$X \to D^{0}  \bar D^{0} \pi^0$ is
\begin{eqnarray}
E_\pi = \frac{ 
\mx^2 - 4 \md^2 + \mpi^2}{ 2 \mx }= 142 \, \mev,   
\end{eqnarray} 
which is just 7 MeV above the $\pi^0$ mass at $134.98\, \mev$. 
The maximum pion momentum is approximately $44\, \mev$, which is comparable
to both the typical $D$-meson momentum, 
$p_D \sim \gamma \lesssim 48 \, \mev$,
and the momentum scale appearing
in the pion-exchange graph, $\mu \simeq 45\, \mev$.  Since the velocity of 
the pions is $v_\pi = p_\pi/m_\pi \leq 0.34$, a non-relativistic treatment 
of the pion fields is valid. In this respect the treatment of pions
differs from ordinary chiral perturbation theory or the $NN$ theory 
of Refs.~\cite{Kaplan:1998tg, Kaplan:1998we}.

The effective Lagrangian includes 
the charm mesons, the anti-charm mesons,  and the pion fields.  
We denote the fields that annihilate the 
$D^{*0}$, $\bar D^{*0}$, $D^{0}$, $\bar D^{0}$,
and $\pi^0$  as $\bd$, $\bdbar$, $D$, $\bar D$, and $\pi$, respectively. 
To the order we are working we will not need diagrams with charged pions and 
charged $D$ mesons so these are neglected in what follows. 
We construct an effective Lagrangian that is relevant for 
low-energy $S$-wave $DD^*$ scattering, 
where the initial and the final states are 
the $C=+$ superposition of 
$D^{0}\bar D^{*0}$ and $D^{*0}\bar D^{0}$:
\begin{eqnarray}
 |D D^*\rangle \equiv \frac{1}{\sqrt{2}}
 \left[
 |D^{0}\bar D^{*0} \rangle + | D^{*0}\bar D^{0} \rangle
 \right]. 
\label{dd*}
\end{eqnarray} 
An interpolating field with these quantum numbers will be used to calculate
the properties of the $X(3872)$.
We integrate out all momentum scales much larger than the momentum 
scale set by $p_D \sim p_\pi \sim \mu$. For $D$ mesons this corresponds 
to kinetic energy $\lesssim 1 \, \mev$,
for pions the kinetic energy is $\lesssim 7 \, \mev$. 
The hyperfine splitting $\Delta$  and $m_\pi$ 
should be treated as large compared to the typical energy scale in the
theory. 
We start from the Lagrangian of 
HH$\chi$PT~\cite{Wise:1992hn,Burdman:1992gh,Yan:1992gz},
which describes the interactions of $D$ and $D^*$ mesons
with Goldstone bosons, and integrate out the scales $m_\pi$ and $\Delta$
by rephasing fields to eliminate the large
components of their energy.  The method is similar to the rephasing used 
to remove the large mass from the energies of the fields in 
heavy-quark effective theory~\cite{Manohar:2000dt}. 
Details are given in Appendix A.
The effective Lagrangian is 
\begin{eqnarray}
 {\cal L}_{ } 
  &=& 
 \bd^{\dagger} \left(i\partial_0 + {\vec{\nabla}^2\over 2 m_{D^*}
		    }\right)\bd 
 +   D^\dagger \left(i\partial_0 + {\vec{\nabla}^2\over 2 m_{D} } \right) D 
\nonumber \\
 && 
 + \bdbar^{\dagger} \left(i\partial_0 + {\vec{\nabla}^2\over 2 m_{D^*}
		    }\right)\bdbar 
 + \bar D^\dagger \left(i\partial_0 + {\vec{\nabla}^2\over 2 m_{D}
		    }\right) \bar D 
 + \pi^\dagger \left(i\partial_0 + {\vec{\nabla}^2\over 2 m_{\pi}}
   + \delta\right) \pi
\nonumber \\
&&+\,  
\left(\frac{g}{\sqrt{2} f_\pi}\right) \,\frac{1}{\sqrt{ 2m_\pi } }
 \left( D \bd^\dagger \cdot \vec{\nabla}\pi  
   + \bar D^\dagger \bdbar \cdot \vec{\nabla}\pi^\dagger \right) + {\rm
 h.c.}
\nonumber \\
 && 
- \,  
\frac{C_0}{2} \, \left(\bdbar D + \bd \bar D \right)^\dagger 
\cdot \left(\bdbar D + \bd \bar D \right) 
\nonumber \\
&&
+  \,   \frac{C_2 }{16} \, 
\left(\bdbar D + \bd \bar D \right)^\dagger 
\cdot \left(\bdbar (\overleftrightarrow \nabla)^2 D 
        + \bd (\overleftrightarrow \nabla)^2 \bar D \right) + h.c.
\nonumber \\
&&+ \,   \frac{B_1 }{\sqrt{2}}\frac{1}{\sqrt{2 m_\pi}} \left(\bdbar D + \bd \bar D \right)^\dagger \cdot D \bar{D} \vec{\nabla} \pi + h.c. + \cdots, 
\label{lag}
\end{eqnarray}
where 
$\delta = \Delta - m_\pi \simeq 7\, \mev$. 
Note that $\mu^2 = \Delta^2 - m_\pi^2 \approx 2 m_\pi \delta $.   
We use the notation 
$\overleftrightarrow \nabla = \overleftarrow\nabla - \overrightarrow
\nabla$, and ``$\cdots$'' 
in Eq.~(\ref{lag}) denotes higher-order interactions.
The pion decay constant is $f_\pi = 132 \,\mev$ with our choice of
normalization.
Notice that, 
since we are only interested in a $C=+$ superposition of the
$D^{0}\bar D^{*0}$ and $D^{*0}\bar D^{0}$
defined in Eq.~(\ref{dd*}), contact interactions 
are written in terms of the combination of fields 
$ \left(\bdbar D + \bd \bar D \right) / \sqrt{2}$. 
Because $\pi$ is a non-relativistic field, $\pi$ annihilates and $\pi^\dagger$
creates $\pi^0$ quanta, so that the Lagrangian in Eq.~(\ref{lag})
allows $ D^{0*}\to D^0 +\pi^0$ and $D^0 + \pi^0\to D^{*0}$ transitions 
and forbids $D^{*0}+\pi^0\to D^0$ and $D^0 \to D^{*0}+\pi^0$. 
Therefore in this effective field theory the only channels that appear
are $\bar{D}^{*0} D^0 + \bar{D}^0 D^{* 0}$ and $D^0 \bar{D}^0 \pi^0$.
In amplitudes with external pions, we must multiply by $\sqrt{2 m_\pi}$ 
because of the normalization of the non-relativistic pion fields.  
In the $X(3872) \to D^0 \bar{D}^0 \pi^0$ decay diagrams, this will cancel the
factors of $1/\sqrt{2 m_\pi}$ in the axial coupling and in the term 
proportional to $B_1$ in Eq.~(\ref{lag}).  

Other channels can of course couple to the $X(3872)$. The three-body channels
$D^\pm \bar{D}^0 \pi^\mp$ and $D^+ D^- \pi^0$ are above the $X(3872)$ 
by only $2.8 \pm 0.6\, \mev$ 
and $3.0 \pm 0.6 \, \mev$, 
respectively. These channels can only appear as virtual intermediate states 
in $X(3872)$ decay  and  self-energy graphs that contain at least 
two pion exchanges. These graphs are NNLO and therefore do not appear at the 
order we are working.~\footnote{This assumes that the interpolating field 
for the $X(3872)$ is $\propto \bdbar D + \bd \bar D$, i.e. 
is constructed from neutral $D$-meson
fields only. Since physical results should not depend on the choice 
of interpolating field, we are 
free to make this choice.} The  $D^{*+} D^-$ threshold lies 
$8.7 \, \mev$ above the $X(3872)$,
and we may integrate out these states because they lie outside the 
range of the effective theory.
If kept in the theory, this intermediate state would also only appear at NNLO.
One may worry about other  nearby thresholds,
especially $J/\psi\, \rho$ and $J/\psi \, \omega$ which are only 
$1.4\pm 1.1 \, \mev$ and $8.2\pm 1.0 \, \mev$, respectively, 
above the $X(3872)$.
The $J/\psi\,\rho$ channel has a much smaller energy gap 
than the others. However, one should take into account that the magnitude
of the complex energy gap includes the width of the $\rho$, 
$\Gamma_\rho/2 = 73 \ {\rm MeV}$, so the $J/\psi\,\rho$ channel can be 
safely integrated out~\cite{Braaten:2005ai}. 
A higher-precision analysis of the $X(3872)$ may
need to include these thresholds explicitly, especially if one wishes 
to describe the decays
$X\to J/\psi \, \pi^+ \pi^-$ and $X\to J/\psi \, \pi^+ \pi^- \pi^0$. 
We leave this to future work.

Matching onto HH$\chi$PT yields the $D^0$, $D^{*0}$, and $\pi^0$ 
kinetic terms as well as the axial $D^{*0}$-$D^0$-$\pi^0$ coupling.  
The coupling constant, $g$, is 
determined from data on the decays of $D^*$ mesons.
The CLEO measurements of the $D^{*+}$ width yields 
$g =  0.59\pm 0.07$ at tree level~\cite{Ahmed:2001xc, Anastassov:2001cw}. 
A NLO analysis of $D^*$ decays in  Ref.~\cite{Stewart:1998ke} yields 
$g=0.27^{+0.06}_{-0.03}$.
A more recent analysis~\cite{Fajfer:2006hi} obtains $g=0.61$ at tree-level
and  $g=0.66\, (0.53)$ at NLO, where the number outside parentheses
refers to the result when virtual low-lying even-parity charmed mesons 
are included in the loop calculations and the number in parentheses refers 
to the result obtained when these states are integrated out. 
The uncertainty in the NLO extraction of $g$ is estimated to be 20\%.  
We will use $g=0.6 \pm 0.1$ in this paper.

The remaining terms in Eq.~(\ref{lag}) with coefficients $C_0$, $C_2$, 
and $B_1$ are contact interactions 
that are not obtained from matching HH$\chi$PT but must also be 
included. They incorporate effects 
that come from shorter distance scales than the scale coming from 
$\pi^0$ exchange.
We have only included operators needed to the order we are working.
$C_0$ and $C_2$ mediate $D^0 \bar{D}^{*0}+\bar{D^0} D^{*0}$ scattering 
in the $C=+$, $S$-wave channel
and have zero and two derivatives, respectively. $B_1$ mediates a transition
between $D^0 \bar{D}^{*0}+\bar{D^0} D^{*0}$ in the $C=+$, $S$-wave channel 
to a state with a $D^0$, $\bar{D}^0$, and $\pi^0$. 

In our power counting, 
$p_D \sim p_{D^*} \sim p_\pi \sim \mu \sim \gamma \sim Q$
and we calculate amplitudes in an expansion in powers of $Q$. 
Since the $D^0$, $D^{*0}$ and 
$\pi^0$ are all non-relativistic, $E_D \sim E_{D^*} \sim E_\pi \sim Q^2$, 
so the propagators
of all particles are order $Q^{-2}$. Loop integrations are order $Q^5$.
The $D^{*0}$-$D^0$-$\pi^0$ axial coupling is order $Q$. 
In the exchange diagram
of Fig.~\ref{pionex}, one can drop the energy dependence in the pion 
propagator.
The factors of $\sqrt{2 m_\pi}$ from the vertices cancel the factors of 
$1/(2 m_\pi)$ in the momentum-dependent term in the pion propagator and 
combine with $\delta$ to give $2 m_\pi \delta = \mu^2$, reproducing 
the expression in Eq.~(\ref{ope}). The pion-exchange amplitude is order
$Q^0$  as is easily seen from Eq.~(\ref{pionex}).

Only counting powers of momentum,
the Feynman rules for the terms in the Lagrangian with coefficients
$C_0$, $C_2$, and $B_1$ are naively of order $Q^0$, $Q^2$ and $Q^1$, 
respectively.
However, with this power counting the theory is perturbative and cannot 
produce a bound state. Instead we will treat $C_0$ non-perturbatively,
along the lines of Ref.~\cite{Kaplan:1998tg,vanKolck:1998bw}, and sum
diagrams with $C_0$ to all orders. At LO, using  the 
power divergence subtraction (PDS) scheme
one then finds~\cite{Kaplan:1998tg}, 
\bea
C_0 = \frac{2 \pi}{M_{DD^*}} \frac{1}{\gamma -\Lambda_{\rm PDS}} \, ,
\eea
where $\Lambda_{\rm PDS}$ is the dimensional-regularization 
parameter.\footnote{The dimensional-regularization 
parameter is usually denoted $\mu$
but we use a different symbol here to avoid confusion with the scale 
appearing in pion exchange.}
Taking $\Lambda_{\rm PDS}$ of order $Q$ we find $C_0$ is order $Q^{-1}$ 
which justifies its resummation. 
In PDS, the coefficient $C_2$ is order $Q^{-2}$ as is  $B_1$, 
as we shall see below.
No other short-distance operators are needed for our NLO calculation 
of $X\to D^0 \bar{D}^0 \pi^0$.
Feynman diagrams with $C_2$ and $B_1$ first contribute to 
$X\to D^0 \bar{D}^0 \pi^0$
at NLO.

In addition to expanding in $Q$, we will make one more approximation in 
the NLO calculation of $X\to D^0 \bar{D}^0 \pi^0$. 
In many cases it greatly simplifies calculations
to expand in $m_\pi/m_D \sim 0.07$. 
This is an approximation we will perform when 
evaluating loop diagrams. It is not systematized in our power counting 
scheme.

As emphasized earlier, the perturbative character of pion exchange
depends on the smallness of the parameter appearing in Eq.~(\ref{Xexp}). 
Our effective theory can be used even
if dimensionless parameters conspire to render pion exchange 
non-perturbative, but in this case 
one-pion exchange would have to be resummed as done in the 
$NN$ system \cite{Nogga:2005hy}.

\section{ Decay rate for $X(3872)\to D^{0} \bar D^{0} \pi^0$ }

Here we describe our method for calculating the width of the $X(3872)$ 
resonance. We consider the 
following two-point function of interpolating fields 
$X^i = (D^0 \bar{D}^{0* i}+\bar{D}^0 D^{0* i})/\sqrt{2}$ 
for the $X(3872)$ with the spin index $i$: 
\bea
G(E)\delta^{ij} = \int d^4x \, e^{-i Et } 
\langle 0 |T[ X^i(x) X^j(0) ] | 0 \rangle 
= i \, \delta^{ij} \frac{Z(-E_X)}{E + E_X + i\Gamma/2} + ... \, ,
\eea
where 
$E_X$ is the binding energy of the $X(3872)$
and the ellipsis represents terms that are less important 
in the resonance region, $E + E_X \sim \Gamma$.
We can define a function $\Sigma(E)$, where $-i \, \Sigma(-E_x)$ 
represents the $C_0$-irreducible graphs contributing to $G(E)$.
Our definition of $\Sigma(E)$ is similar to the function $\Sigma$ 
defined in Appendix A of Ref.~\cite{Kaplan:1998sz}.
In terms of $\Sigma(E)$, $G(E)$ is 
\bea
G(E) &=& \frac{-i \, \Sigma(E)}{1 +  C_0 \, \Sigma(E)} \nn \\
&=& \frac{-i \,{\rm Re}\, \Sigma(E) + {\rm Im}\, \Sigma(E)}
{1+ C_0 {\rm Re}\, \Sigma(E) + i \, C_0 \, {\rm Im}\,\Sigma(E)}\, .
\eea
Since the real part of the denominator must vanish at $E = - E_X$, 
we have $1 +  C_0 \, {\rm Re}\,\Sigma(-E_X) = 0$, and 
expanding about $E=-E_X$ we obtain for $G(E)$ 
\bea\label{gexp}
G(E) &=&  \frac{i ( 1/C_0 -(E+E_X)  {\rm Re} \, \Sigma^\prime(-E_X)) 
+  {\rm Im} \,\Sigma(-E_X)}{C_0 (E+E_X) {\rm Re} 
\, \Sigma^\prime(-E_X) + i \,C_0 {\rm Im} \,\Sigma(-E_X)} \nn \\
&=&  \frac{i}{C_0^2 (E+E_X) {\rm Re} 
\, \Sigma^\prime(-E_X) + i \, C_0^2 \,{\rm Im} \,\Sigma(-E_X)} 
- \frac{i}{C_0} \, ,
\eea
where $\Sigma^\prime = d\Sigma/dE$. From Eq.~(\ref{gexp}), 
we immediately see that
\bea
Z(E) = \frac{1}{C_0^2 \,{\rm Re} \, \Sigma^\prime(-E_X)}\, , 
\quad \quad 
\Gamma =\frac{2 \,{\rm Im} \,\Sigma(-E_X)}{{\rm Re}\,\Sigma^\prime(-E_X)} \, .
\eea
The function $2\, {\rm Im}\,\Sigma$ corresponds to the square of the 
decay diagrams. It is interesting to compare the result
of evaluating the loop diagrams and taking the real part with direct 
evaluation of the decay diagrams.

Consider for example the evaluation of the two-loop diagram in 
Fig.~\ref{sigmaNLO}a in Appendix B.
The result of evaluating the  graph is 
\bea
{\rm Fig}.~\ref{sigmaNLO}{\rm a})&=& - i \frac{g^2}{2f_\pi^2}\frac{1}{2 m_\pi}
\left(\frac{\Lambda_{\rm PDS}}{2}\right)^{8 - 2 D}\\
&&\times \int \frac{d^Dq}{(2\pi)^D} \int \frac{d^Dl}{(2\pi)^D} \, 
\frac{1}{q_0+E_X/2 -q^2/(2 m_{D^*} )
+ i\epsilon} \, \frac{1}{-q_0+E_X/2 -q^2/(2 m_{D} )+ i\epsilon} \nn \\ 
&&\times \frac{1}{l_0+E_X/2 -l^2/(2 m_{D^*}) + i\epsilon}\, 
\frac{1}{-l_0+E_X/2 -l^2/(2 m_{D} )+ i\epsilon} \nn \\
&&\times \frac{(q+l)_i (q+l)_j}{q_0 + l_0 -(q+l)^2/(2m_\pi) + \delta 
+ i \epsilon} \nn \, .
\eea
We perform the energy integrals by contour integration, taking 
the poles of the $D$-meson propagators.
This yields
\bea\label{7a}
{\rm Fig}.~\ref{sigmaNLO}{\rm a})&=&i \frac{g^2}{2f_\pi^2}
\left(\frac{\Lambda_{\rm PDS}}{2}\right)^{8 - 2 D} \nn \\
&& \times
\int \frac{d^{D-1}q}{(2\pi)^{D-1}} \int \frac{d^{D-1}l}{(2\pi)^{D-1}} \, 
\frac{1}{E_X -q^2/(2 M_{DD^*} )+ i\epsilon} \, 
\frac{1}{E_X -l^2/(2 M_{DD^*} )+ i\epsilon} \nn \\  
&&\times \quad \frac{(q+l)_i (q+l)_j}
                    {2 m_\pi(E_X -q^2/(2 m_D)-l^2/(2 m_D)) -(q+l)^2 
+ \mu^2+ i \epsilon} \nn \\
\nn \\
&=& i  \frac{g^2}{2f_\pi^2}(2 M_{DD^*})^2 \left(\frac{\Lambda_{\rm PDS}}{2}
\right)^{8 - 2 D} \nn \\
&&\times \int \frac{d^{D-1}q}{(2\pi)^{D-1}} 
\int \frac{d^{D-1}l}{(2\pi)^{D-1}} \, 
\frac{1}{q^2 + \gamma^2 - i\epsilon}\, \frac{1}{l^2 + \gamma^2 - i\epsilon}  
\nn \\
&&\times \quad\frac{(q+l)_i (q+l)_j}
                   {- m_\pi(\gamma^2/M_{DD^*} +q^2/m_D+l^2/m_D)) -(q+l)^2 
+\mu^2 +i \epsilon} \, .
\eea
The first two propagators that come from the $D^*$ mesons clearly scale 
as $Q^{-2}$. The 
last two terms in the pion propagator denominator, $-(q+l)^2 + \mu^2$, 
scale as $Q^2$ while
the other terms scale as $(m_\pi/m_D) Q^2$. Since $m_\pi/m_D \sim 0.07$ 
is comparable to our expansion
parameter in Eq.~(\ref{Xexp}),
these terms can be systematically dropped. The neglected terms come 
from the pion kinetic energy, and in dropping them we are treating 
the pions in the potential approximation~\cite{Mehen:1999hz}.
The final answer is then
\bea
{\rm Fig}.~\ref{sigmaNLO}{\rm a}) &=&-i \frac{g^2}{2f_\pi^2}(2 M_{DD^*})^2 
\left(\frac{\Lambda_{\rm PDS}}{2}\right)^{8 - 2 D}
 \nn \\
&&\times \int \frac{d^{D-1}q}{(2\pi)^{D-1}} 
\int \frac{d^{D-1}l}{(2\pi)^{D-1}} \, 
\frac{1}{q^2 + \gamma^2 - i\epsilon} \,\frac{1}{l^2 + \gamma^2 -i\epsilon}  
\,\frac{(q+l)_i (q+l)_j}{(q+l)^2 - \mu^2 - i \epsilon} \nn \\
&=& -i\frac{g^2}{2f_\pi^2}\frac{\delta_{ij}}{3}
\left(\frac{M_{DD^*}}{2\pi}\right)^2
\left[ (\Lambda_{\rm PDS} -\gamma)^2 
+\mu^2\left(\frac{1}{4\hat \epsilon} +\frac{1}{2} 
+ \log \left(\frac{\Lambda_{\rm PDS}}{2 \gamma -i \mu}
\right)\right)\right] \, .
\eea
where $1/(4\hat \epsilon)= 1/(4 \epsilon) + ({\rm ln}\, \pi -
\gamma_{\rm E})/2$.
We have used three-dimensional rotational invariance to replace
$(q+l)_i (q+l)_j$ with $(q+l)^2 \delta_{ij}/3$ and use the PDS scheme 
to evaluate the remaining scalar integrals.

There is one instance when dropping $m_\pi/m_D$ corrections is not 
appropriate. 
To see this consider evaluating the real part of Fig.~\ref{sigmaNLO}a 
by evaluating the cut 
diagram. The cut runs through the $D$-meson and pion propagators. 
In the cut diagrams,
these propagators are replaced with $\delta$-functions. 
For the $D$-meson propagators
integrating over the $\delta$-functions is equivalent to taking 
the pole using contour integration.
So the cut diagram is obtained from Eq.~(\ref{7a}) simply by replacing 
the pion propagator 
with the corresponding $\delta$-function. 
Doing this and making the substitutions
$q\to p_{\bar{D}}$ and $l \to p_D$ so that $q+l= p_D + p_{\bar D}= -p_\pi$, 
one obtains
\bea\label{cut7a}
&&\frac{g^2}{2 f_\pi^2}(2 M_{DD^*})^2 
\int \frac{d^3 p_D d^3p_{\bar{D}} }{(2\pi)^5} \nn \\
&&\times |\vec{\epsilon} \cdot \vec{p}_\pi|^2 
\frac{1}{p_D^2+\gamma^2} \frac{1}{p_{\bar D}^2+\gamma^2}
\; \delta\left(\mu^2- p_\pi^2 -\frac{m_\pi}{ M_{DD^*}} \gamma^2 
-  \frac{m_\pi}{m_D} p_D^2
 -  \frac{m_\pi}{m_D} p_{\bar D}^2 \right) \, .
\eea
This clearly reproduces the interference term in Voloshin's effective-range 
calculation of 
$X\to D^0 \bar{D}^0 \pi^0$ \cite{Voloshin:2003nt}. 
The $\delta$-function in Eq.~(\ref{cut7a}) 
imposes the 
constraint on the phase space due to energy conservation. 
Dropping $m_\pi/m_D$-suppressed 
terms in the $\delta$-function corresponds to neglecting the final-state 
$D$-meson's kinetic energy
and would leave the integrals over their momentum unconstrained. 
Clearly this is not a good approximation. 
Physically, it is also clear that the on-shell propagating pion in the final
state cannot be treated in the potential approximation.

Therefore, in evaluating  ${\rm Im}\, \Sigma(-E_X)$ we will calculate 
the decay amplitudes for the diagrams and integrate
over the physical three-body phase. In diagrams with virtual pions, 
we drop the kinetic energy so the pions are potential.\footnote{For 
further discussion on the role
of recoil corrections, see Ref. \cite{Baru:2004kw}.}
In the virtual diagrams
this approximation is valid up to $O(m_\pi/m_D)$ corrections. 
Since our expansion parameter is expected
to be $0.05-0.1$, making this approximation in the virtual NLO graphs 
induces an error of the
same size as the NNLO correction.

\begin{figure}[t]
\begin{center}
\includegraphics[width=3cm]{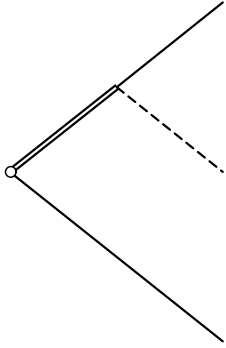}
\caption{ LO diagram for decay rate. \label{lodecay}}
\end{center}
\end{figure}

\begin{figure}[t]
\begin{center}
\includegraphics[width=3cm]{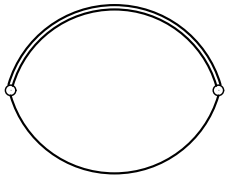}
\caption{LO diagram for calculating wavefunction renormalization.
\label{sigmaLO}}
\end{center}
\end{figure}

The LO decay diagram is shown in Fig.~\ref{lodecay}. 
The $D^*$ propagator scales as 
$1/Q^2$ and the axial coupling scales as $Q$ so the LO diagram 
is order $Q^{-1}$. We show only 
one diagram, but there are two channels related by $C$-conjugation 
that are implied.
It is straightforward to evaluate these diagrams and obtain
\bea
i \frac{g }{f_\pi} \, \frac{M_{DD^*}}{p_D^2 +\gamma^2}\,\vec{p}_\pi 
\cdot \vec{\epsilon}_X + (p_D \to p_{\bar D}) \, ,
\eea
The LO contribution to the wavefunction diagram is shown 
in Fig.~\ref{sigmaLO}.
The graph is $O(Q)$ and therefore ${\rm Re}\, \Sigma^\prime(-E_X)$ 
is $O(Q^{-1})$. 
The result
of evaluating this graph and taking the derivative is 
\bea
{\rm Re}\, \Sigma^\prime_{LO} = \frac{M_{DD^*}^2}{2 \pi \gamma}\, .
\eea
The LO decay diagram in Fig.~\ref{lodecay} is $O(Q^{-1})$ so the 
leading contribution to 
${\rm Im}\, \Sigma(-E_X)$
from Fig.~\ref{lodecay} is $O(Q^{-2})$. 
Dividing by the LO wavefunction renormalization 
which is $O(Q^{-1})$
one sees that the leading contribution to the decay rate is $O(Q^{-1})$. 
The result 
reproduces Voloshin's calculation of 
$X\to D^0 {\bar D}^0 \pi^0$~\cite{Voloshin:2003nt}:
\bea
\frac{d \Gamma_{\rm LO}}{d p_D^2 d p_{\bar D}^2}= 
\frac{g^2}{32 \pi^3 f_\pi^2} 2 \pi \gamma (\vec{p}_\pi 
\cdot \vec{\epsilon}_X )^2
\bigg[ \frac{1}{p_D^2 + \gamma^2} + \frac{1}{p_{\bar D}^2 + \gamma^2}\bigg]^2
\eea

\begin{figure}[t]
\begin{center}
a) \includegraphics[width=3cm]{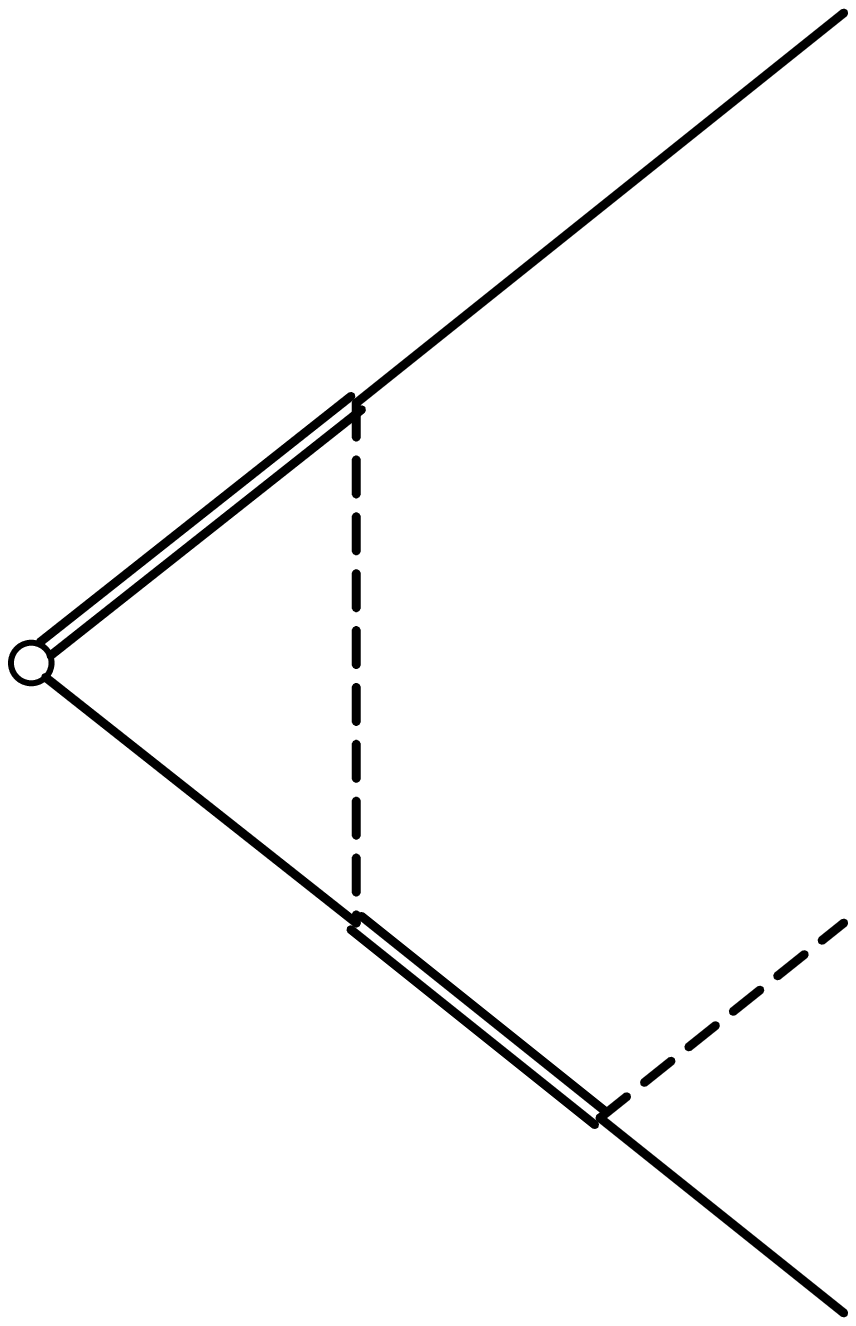} \hspace{1.0 in}
b) \includegraphics[width=3cm]{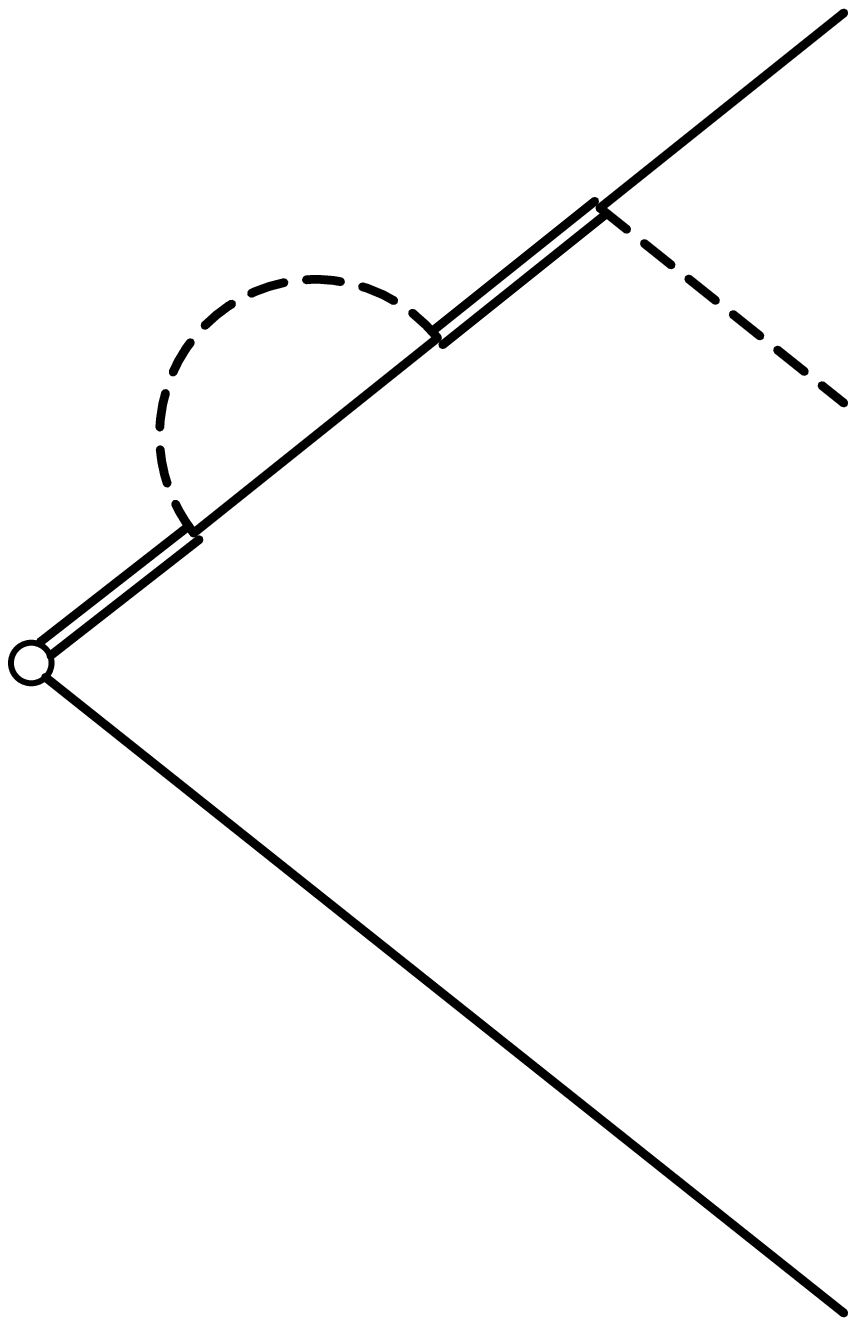}
\caption{ NLO diagrams for the decay rate involving pion exchange. 
\label{nlopiondecay}}
\end{center}
\end{figure}

\begin{figure}[t]
\begin{center}
a) \includegraphics[width=4cm]{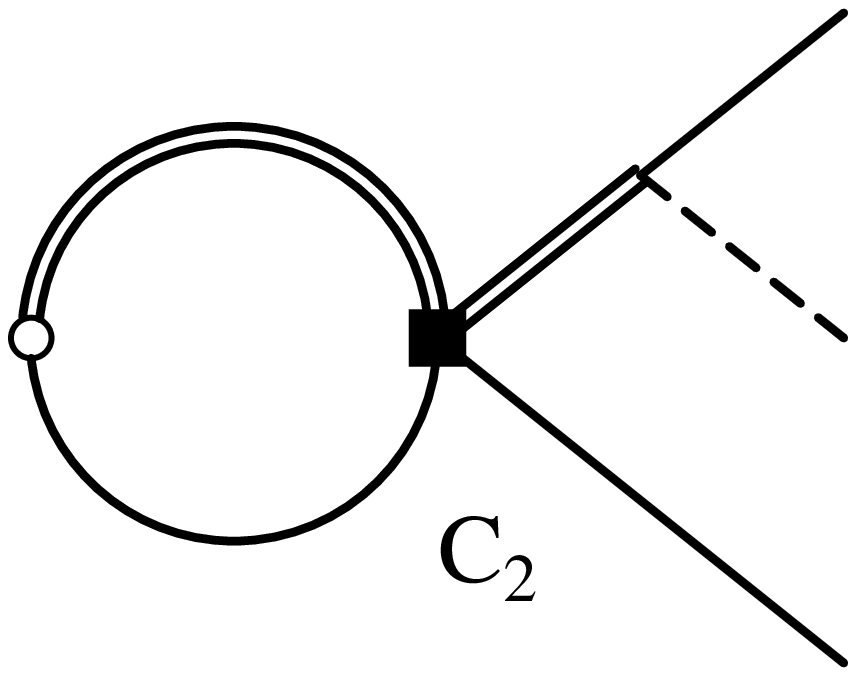} \hspace{1.0 in}
b) \includegraphics[width=4cm]{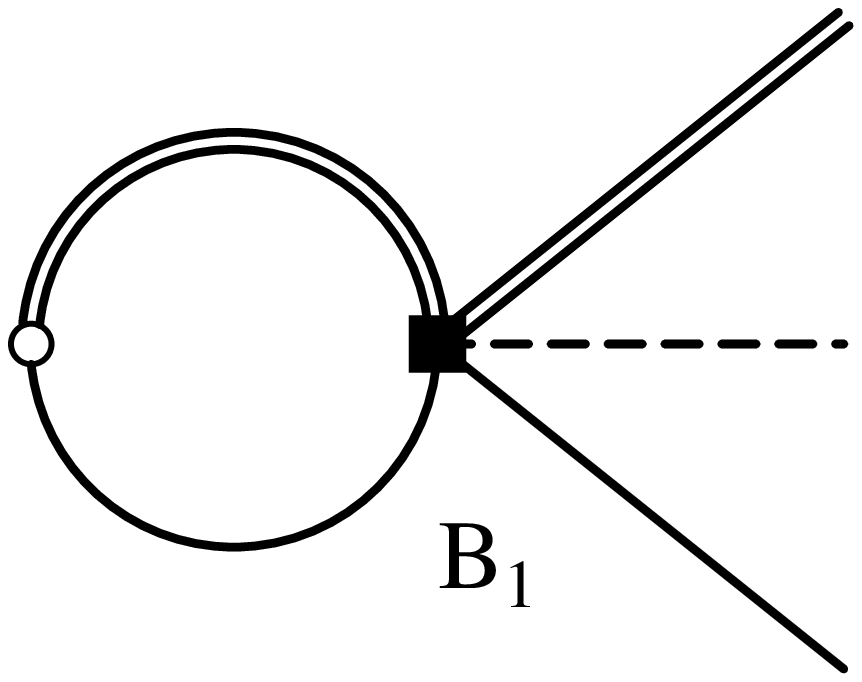}
\caption{ NLO diagrams for the decay rate which involve contact interaction. 
\label{nlocontactdecay}}
\end{center}
\end{figure}

The NLO corrections to the decay rate are suppressed by one power of $Q$.
They come from graphs, shown in 
Figs.~\ref{nlopiondecay} and \ref{nlocontactdecay}, 
with one additional pion exchange 
or one insertion of $C_2$ and $B_1$. These coefficients scale as $Q^{-2}$.
NLO contributions to the wavefunction renormalization are down by one power 
of $Q$ as well.
These contributions are given by the two-loop self-energy  diagrams involving 
pion exchange 
or an insertion of $C_2$ shown in Fig.~\ref{sigmaNLO}.

\begin{figure}[t]
\begin{center}
a) \includegraphics[width=3cm]{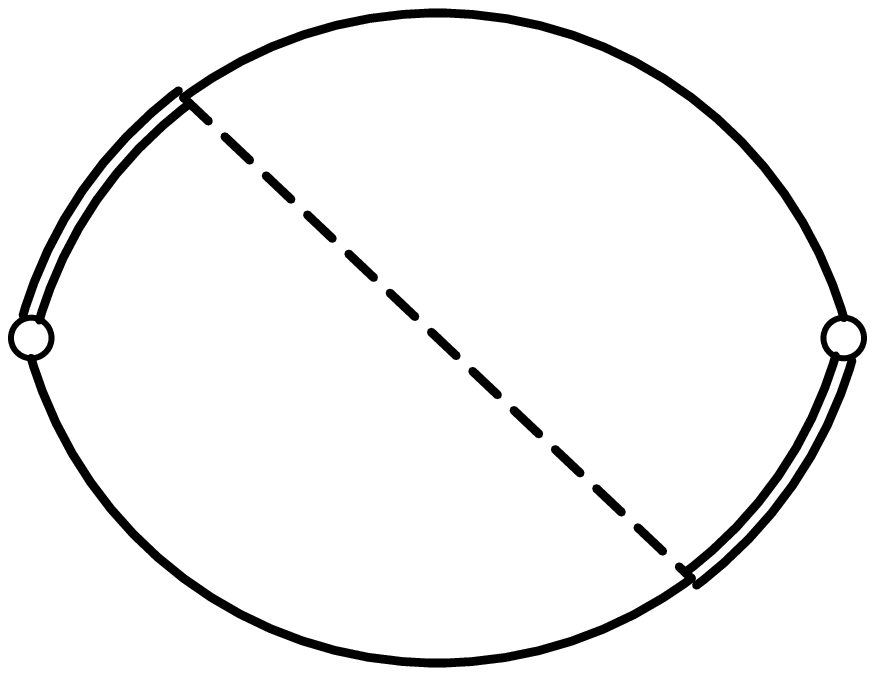} \hspace{0.5 in}
b) \includegraphics[width=3cm]{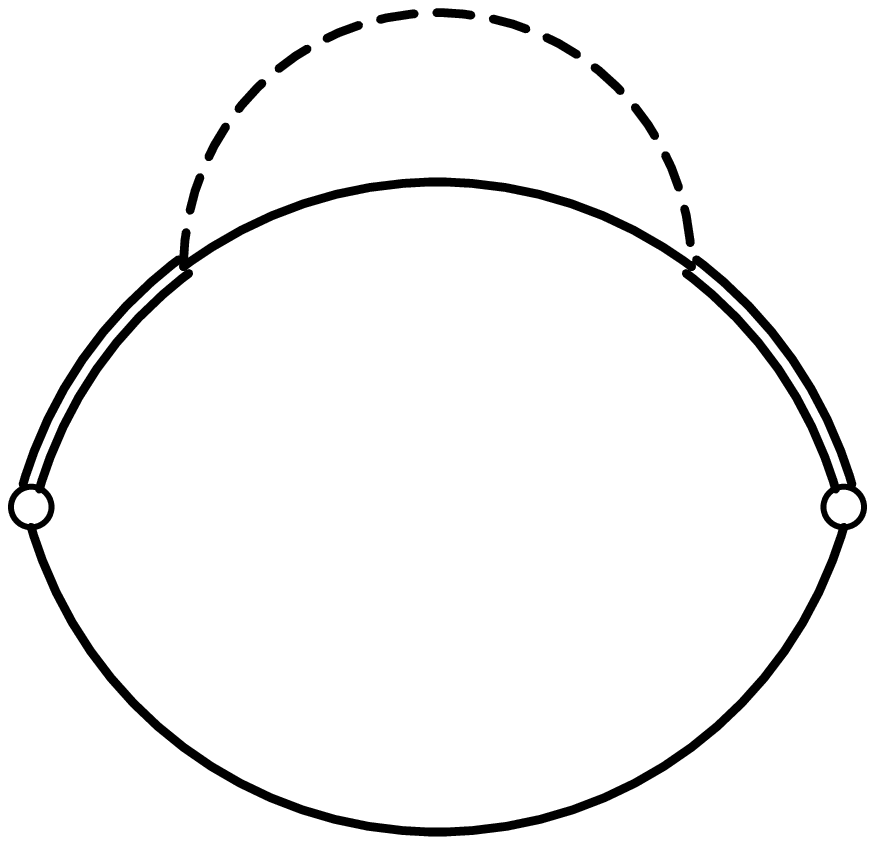} \hspace{0.5 in}
c) \includegraphics[width=4cm]{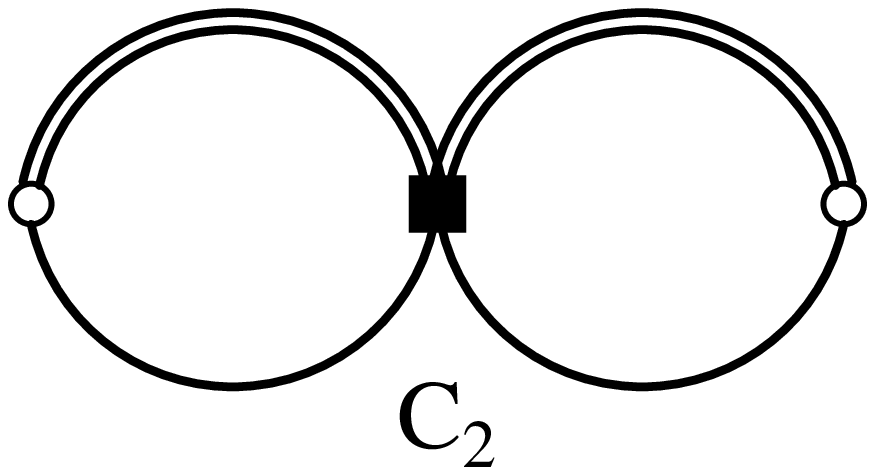} 
\caption{NLO diagrams for calculating wavefunction renormalization.
\label{sigmaNLO}}
\end{center}
\end{figure}

The results for individual diagrams are given in Appendix B. 
The final expression
for the NLO differential rate is 
\bea\label{NLOrate}
\frac{d \Gamma_{\rm NLO}}{d p_D^2 d p_{\bar D}^2} &=& 
\frac{d \Gamma_{\rm LO}}{d p_D^2 d p_{\bar D}^2}\bigg(1 
+ \frac{g^2 M_{DD^*} \gamma}{6 \pi f_\pi^2}
\left(\frac{4 \gamma^2-\mu^2}{4 \gamma^2+\mu^2}\right)
+ C_2(\Lambda_{\rm PDS}) 
\frac{M_{DD^*} \gamma (\gamma-\Lambda_{\rm PDS})^2}{\pi} \bigg) \\
&&\hspace{-0.4 in}-\frac{g \gamma}{16 \pi^3 f_\pi}
\bigg( \frac{g M_{DD^*}}{f_\pi} C_2(\Lambda_{\rm PDS})  
+ B_1(\Lambda_{\rm PDS})\bigg)(\Lambda_{\rm PDS}-\gamma)\, 
(\vec{p}_\pi \cdot \vec{\epsilon}_X )^2 \, 
\bigg[ \frac{1}{p_D^2 + \gamma^2} + \frac{1}{p_{\bar D}^2 + \gamma^2}\bigg]
\nn \\
&&\hspace{-0.4 in}- \frac{g^4 M_{DD^*} \gamma}{64 \pi^3 f_\pi^4}  
(\vec{p}_\pi \cdot \vec{\epsilon}_X )^2 
\bigg[\frac{{\rm Re}\, h_1(p_D)}{p_D^2 + \gamma^2} 
+ \frac{{\rm Re} \,h_1(p_{\bar D})}{p_{\bar D}^2 + \gamma^2}\bigg]
\bigg[ \frac{1}{p_D^2 + \gamma^2} + \frac{1}{p_{\bar D}^2 + \gamma^2}\bigg] 
\nn \\
&&\hspace{-0.4 in}+ \frac{g^4 M_{DD^*}\gamma}{64 \pi^3 f_\pi^4} 
\bigg[\frac{{\rm Re}\, h_2(p_D)}{p_D^2 + \gamma^2}
\,\vec{p}_\pi 
\cdot \vec{\epsilon}_X\,\vec{p}_D \cdot \vec{\epsilon}_X 
\,\vec{p}_\pi \cdot \vec{p}_D + (p_D \to p_{\bar D})\bigg]
\bigg[ \frac{1}{p_D^2 + \gamma^2} + \frac{1}{p_{\bar D}^2 + \gamma^2}\bigg]  
\, . \nn
\eea
The functions $h_1(p)$ and $h_2(p)$ are given in Appendix B. 
The first line in Eq.~(\ref{NLOrate}) 
is a multiplicative correction to the LO decay rate. 
Note that in the absence of pions
\bea\label{c2}
C_2(\Lambda_{\rm PDS}) = \frac{2 \pi}{M_{DD^*}} \frac{r_0}{2}
\frac{1}{(\Lambda_{\rm PDS} -\gamma)^2} \, ,
\eea
where $r_0$ is the effective range. 
The term proportional to $C_2$ in the first line 
of Eq.~(\ref{NLOrate}) reproduces the expected correction from the 
effective-range 
theory, in which the leading correction involving $r_0$ comes 
from the modification of the 
normalization of the wavefunction:
\bea
\psi^{\rm ER}(r) = 
\sqrt{\frac{\gamma}{4 \pi (1-\gamma r_0)}}\frac{e^{-\gamma r}}{r} \, .
\eea
The second line in Eq.~(\ref{NLOrate}) is the interference between 
a short-distance
local coupling of the $X$ to the $D^0 \bar{D}^0 \pi^0$ state and the LO 
amplitude. Note the coefficient of this term scale as 
$1/(\Lambda_{\rm PDS} -\gamma)$
and disappears if one takes $\Lambda_{\rm PDS} \to \infty$, confirming 
the short-distance nature of the contribution. 
The final terms are non-analytic corrections due to pion exchange.
These contributions turn out to give a very small ($\sim 1\%$) 
contribution to the 
decay rate, so the NLO correction is entirely dominated 
by the contact interaction contributions.

\begin{figure}[t]
\begin{center}
\includegraphics[width=12cm,angle=270]{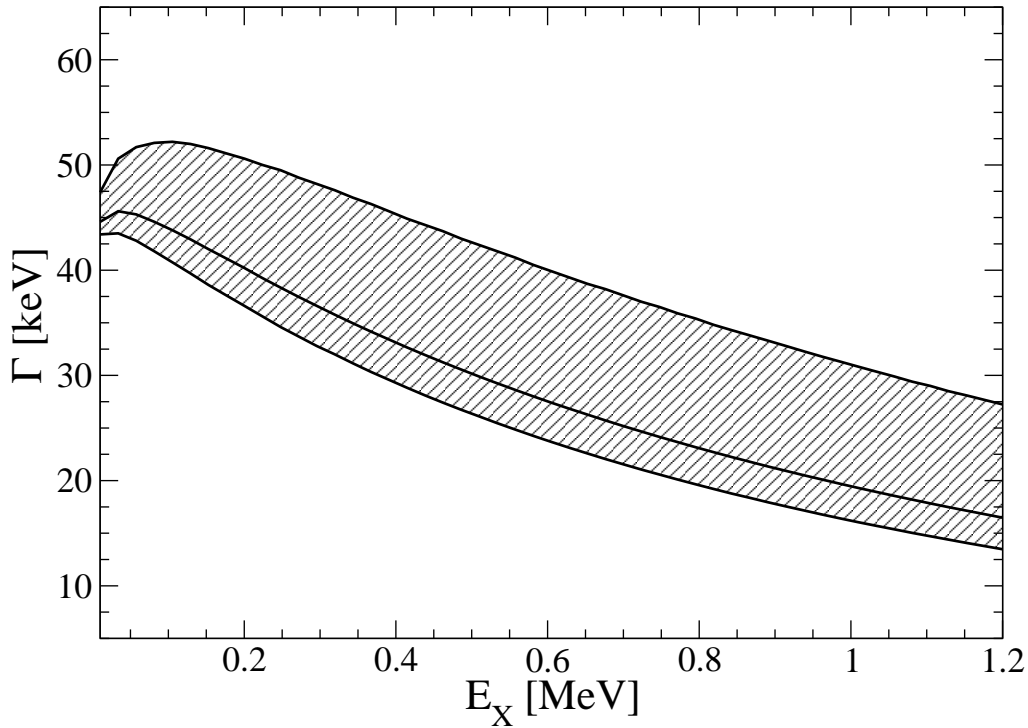}
\caption{ Decay rate for $X\to D^{0}\bar D^{0}\pi^0$ as a 
function of $E_X$. We use $g=0.6$. The central solid line 
corresponds to the 
LO prediction. The band is the 
result of the NLO calculation when the parameters $r_0$ and $\eta$
are varied in the ranges
$0 \leq r_0 \leq (100 \,{\rm MeV})^{-1}$ and $-1 \leq \eta \leq 1$.
\label{x-ddp}}
\end{center}
\end{figure}

We will parametrize $C_2$ according to Eq.~(\ref{c2}), 
where $r_0$ is to be interpreted
as the short-distance contribution to the effective range. 
Since we have integrated out the scales $m_\pi$ and $\Delta$, 
it is reasonable to take $r_0 \sim (100 \, {\rm MeV})^{-1}$.
We will parametrize 
\bea
\bigg( \frac{g M_{DD^*}}{f_\pi} C_2(\Lambda_{\rm PDS}) 
+ B_1(\Lambda_{\rm PDS})\bigg)(\Lambda_{\rm
PDS}-\gamma) = \frac{\eta}{(100\, {\rm MeV})^3} \, , 
\eea
where $\eta$ is a dimensionless parameter we expect to be of order unity.
Fig.~\ref{x-ddp} shows the  partial width $\Gamma[X\to D^0 \bar{D}^0 \pi^0]$ 
as a function of the binding energy. 
The central solid line is the LO result. We use the central 
value for the tree-level extraction of the $D$-meson axial coupling, 
$g = 0.6$. 
The band in Fig.~\ref{x-ddp} shows the NLO rate with the  parameters $r_0$ 
and $\eta$ varied between
\bea
0 \leq r_0 \leq \frac{1}{100\, {\rm MeV}}\, , \qquad -1 \leq \eta \leq 1 \, . 
\eea
As stated earlier the non-analytic 
calculable corrections from pion exchange in Eq.~(\ref{NLOrate}) 
give negligible corrections. The band is 
dominated entirely by the contact interaction contributions.
Measurements of the $X$ mass and partial decay width into
$D^{0}\bar{D}^{0}\pi^{0}$ can naturally be explained within 
a molecular picture if the corresponding point in Fig.~\ref{x-ddp} 
falls within, 
or ---due to higher orders--- close to, 
this band. Values far outside the band can be acommodated only if
short-range parameters or higher-order effects are anomalously large. 
In either case the appeal of our framework would be strongly diminished.

\section{Summary}

In this paper we have developed an effective field theory of non-relativistic 
pions and $D$ mesons that can be used to describe the properties of 
the $X(3872)$, assuming it is a weakly bound state of $D^0 \bar{D}^{*0}$ 
and $D^{*0} \bar{D}^{0}$ with anomalously small binding energy. 
Because of an accidental cancellation between 
the $D$-meson hyperfine splitting and the mass of the $\pi^0$, 
pion exchange is characterized by a smaller scale than is typically 
the case in nuclear physics. 
This relatively small scale and the small axial coupling 
in the $D$-meson system (compared to the nucleon's axial coupling) 
combine to make the corrections from $\pi^0$-meson exchange amenable 
to perturbation theory. 
This justifies the application of a theory similar to that
proposed by Kaplan, Savage, and Wise for low-energy $NN$ interactions
\cite{Kaplan:1998tg,Kaplan:1998we}, 
in which a leading-order contact interaction is resummed to all orders 
to produce the bound state, and pion exchange and higher-derivative 
contact interactions are treated within perturbation theory.

This theory reproduces at leading order the calculation of 
$\Gamma[X\to D^0 \bar{D}^0 \pi^0]$ by Voloshin \cite{Voloshin:2003nt}
which exploits the universal behavior of the $D D^*$ wavefunction in limit 
of small binding energy. 
Effective-range corrections as well as other corrections from 
short-distance scales are encoded in higher-dimension contact operators 
in the theory. These corrections turn out to completely 
dominate non-analytic calculable corrections from $\pi^0$ exchange. 
Varying these coefficients within 
ranges determined  by naturalness allows us to estimate the size of 
corrections to the leading-order calculations of Voloshin. 
While it is somewhat disappointing that the non-analytic calculable 
corrections from $\pi^0$  exchange are so small that an experimental 
test of this aspect of the theory 
seems unlikely in the foreseeable future, 
the smallness of these corrections confirms one of the main points of 
this work, 
namely that pion exchange can be dealt with using perturbation theory.

A naive estimate of the size of the NNLO corrections based on the expansion 
parameter in 
Eq.~(\ref{Xexp}) is 1\% or smaller. 
It is important to remember that in conventional 
nuclear physics, large corrections come from graphs with two or more pion 
exchanges in 
the $^3S_1$ channel, which first arise at NNLO. 
The two-pion exchange graphs at NNLO graphs come with
large coefficients, $\sim 5$, which ruin the perturbative expansion of
KSW for two-nucleon systems~\cite{Fleming:1999ee}. 
In our case similar size coefficients in two-pion exchange 
graphs should not ruin perturbation theory since even with a large 
coefficient $\sim 5$, 
they would only be expected to be
 5\% or smaller. 
It would be interesting to perform the NNLO calculation to check this. A 
NNLO correction of 5\% would dominate the non-analytic NLO contribution
but would be smaller than the uncertainty in the contact interaction 
contribution, indicating 
convergence of the expansion.
In the unlikely case that pion exchange is non-perturbative,
it can be resummed as done in nuclear physics \cite{Nogga:2005hy}.

It is straightforward to extend the analysis of this paper to other 
$X(3872)$ decay and production
processes, such as $X\to D^0 \bar{D}^0 \gamma$ 
or $X\to J/\psi \rho^* \to J/\psi \pi^+ \pi^-$. 
Coupling to $J/\psi \, \rho$ and $J/\psi \,\omega$ channels
can be incorporated by including these degrees of freedom explicitly 
in the theory and coupling them to
$D^0  \bar{D}^{*0} + D^{*0}  \bar{D}^{0}$  via contact interactions. 
It would be interesting to calculate $\pi^0$ 
exchange to other decays of the $X(3872)$ to see if these corrections 
lead to any interesting observable effects.
It would also be interesting to use data or theoretical calculations 
to fix some of the counterterms appearing in the
theory so as render calculations in this paper more predictive. 

\acknowledgments 

We would like to thank E. Braaten and C. Hanhart for useful discussions.
This work was supported in part by the U.S. Department of Energy under 
grant numbers
DE-FG02-06ER41449 (S.F. and M.K.), 
DE-FG02-05ER41368 (T.M.), DE-FG02-05ER41376 (T.M.), 
DE-AC05-84ER40150 (T.M.), and DE-FG02-04ER41338 (U.v.K.),
by the Nederlandse Organisatie voor Wetenschappelijk Onderzoek  (U.v.K.),
and by Brazil's FAPESP under a Visiting Professor grant (U.v.K.).
We would also 
like to thank the hospitality of the University of Arizona (T.M.),
the Center for Theoretical Physics at MIT (T.M.),
the Kernfysisch Versneller Instituut at Rijksuniversiteit Groningen (U.v.K.),
the Instituto de F\'\i sica Te\'orica of the
Universidade Estadual Paulista (U.v.K.),
and the Instituto de F\'\i sica of the Universidade de S\~ao Paulo (U.v.K.)
where part of this work was completed.

\appendix
\section{Deriving the EFT Lagrangian from HH$\chi$PT}\label{hhchipt}

Heavy-hadron chiral perturbation theory 
(HH$\chi$PT )~\cite{Wise:1992hn,Burdman:1992gh,Yan:1992gz} can be 
used to derive the low-energy effective Lagrangian for $D$ mesons, 
$D^{*}$ mesons, and
pions relevant to the $X(3872)$. We begin with the  
two-component HH$\chi$PT Lagrangian introduced in Ref.~\cite{Hu:2005gf},
\begin{eqnarray}
 {\cal L}  =
{\rm Tr}\left[ H^\dagger (i D_0) H \right] 
  - g \, {\rm Tr}\left[ H^\dagger H \bsigma\cdot {\bf A} \right] 
 + {\Delta_{} \over 4} \,
{\rm Tr}\left[ H^\dagger \bsigma H\bsigma\right], 
\label{append-L}
\end{eqnarray}
where $\bsigma$ are the Pauli matrices, $\Delta_{}=m_{H^*}-m_{H}$, 
$H = {\bf D}\cdot\bsigma+ D$,
and ${\bf A}= -\vec{\nabla}\pi/f_\pi +{\cal O}(\pi^3) $. Here
${\bf D}$ is a heavy vector field, $D$ is a heavy pseudoscalar field,
and $\pi$ is the pion field,
\begin{eqnarray}
 \pi = \begin{pmatrix} {1\over \sqrt{2}}\pi^0 & \pi^+ \\ 
\pi^- & -{1\over \sqrt{2}}\pi^0   \end{pmatrix}.
\end{eqnarray}
Evaluating the traces in Eq.~(\ref{append-L}) we obtain
\begin{eqnarray}
 {\cal L} &=& 
2{\bf D}^\dagger \left(iD_0-{\Delta_{} \over 4} \right) {\bf D} 
+ 2 D^\dagger \left(iD_0 + {3\Delta_{} \over 4} \right) D  \nonumber
\\
&&\hspace{2cm}
- 2 g \, \left(
{\bf D}^\dagger \cdot {\bf A} D + D^\dagger {\bf D}\cdot {\bf A}
\right) 
-2ig\, {\bf D}^\dagger\cdot {\bf D}\times {\bf A} \,.
\label{append-L2}
\end{eqnarray}
Since we wish to describe a bound state of two heavy mesons the power counting
of HH$\chi$PT in powers of $1/m_H$ is inappropriate. Instead we need to power
count in the relative velocity $v \ll 1$ of the heavy mesons. The
kinetic energy which is sub-leading in $1/m_H$ is leading in $v$, 
and as a consequence we must include the kinetic term in our Lagrangian,
\begin{eqnarray}
 {\cal L} &=& 
2{\bf D}^\dagger \left(iD_0+ \frac{\vec{\nabla}^2}{2m_{D^*}}-\frac{\Delta}{4} 
\right) {\bf D} 
+ 2 D^\dagger \left(iD_0 + \frac{\vec{\nabla}^2}{2m_{D}}+ {3\Delta_{} \over 4}
\right) D  \nonumber
\\
&&\hspace{2cm}
- 2 g \, \left(
{\bf D}^\dagger \cdot {\bf A} D + D^\dagger {\bf D}\cdot {\bf A}
\right) 
-2ig\, {\bf D}^\dagger\cdot {\bf D}\times {\bf A} \,.
\label{append-L2.5}
\end{eqnarray}
We now rescale the heavy-meson fields
\begin{eqnarray}
 \{D, {\bf D}\} \to \frac{1}{\sqrt{2}} \;e^{i 3\Delta t/4} \{D, {\bf D} \} \,,
\end{eqnarray}
which gives 
\begin{eqnarray}
 {\cal L} &=& 
{\bf D}^\dagger \left(iD_0+\frac{\vec{\nabla}^2}{2m_{D^*}}-\Delta_{} 
\right) {\bf D} 
+D^\dagger \left(iD_0+\frac{\vec{\nabla}^2}{2m_{D}} \right) D  \nonumber
\\ 
&&\hspace{2cm}
- g \, \left(
{\bf D}^\dagger \cdot {\bf A} D + D^\dagger {\bf D}\cdot {\bf A}
\right) 
- ig\, {\bf D}^\dagger\cdot {\bf D}\times {\bf A} \,.
\label{append-L3}
\end{eqnarray}
Since we are only interested in those terms involving $D^{*0}$, $D^{0}$, 
and $\pi^0$ 
we keep only these fields in the Lagrangian,
\begin{eqnarray}
 {\cal L} &=& 
{\bf D}^\dagger \left(iD_0+ \frac{\vec{\nabla}^2}{2m_{D^*}}-\Delta_{} 
\right) {\bf D} 
+D^\dagger \left(iD_0 + \frac{\vec{\nabla}^2}{2m_{D}}\right) D  \nonumber
\\ 
&&\hspace{2cm}
+ \frac{g}{\sqrt{2}f_\pi} \, \left(
{\bf D}^\dagger \cdot\vec{\nabla}\pi^0 D + D^\dagger {\bf D}\cdot 
\vec{\nabla}\pi^0 \right) 
-i\frac{g}{\sqrt{2}f_\pi}\, {\bf D}^\dagger\cdot {\bm D}\times 
\vec{\nabla}\pi^0.
\label{append-L4}
\end{eqnarray}

Next the kinetic term for the pion is derived from the chiral Lagrangian
\begin{eqnarray}
 {\cal L}_{\pi} 
&=&
 \frac{f^2_\pi}{8} {\rm Tr}\left[\partial_\mu \Sigma \partial^\mu
       \Sigma^\dagger \right] +\frac{f^2_\pi}{4} B_0 {\rm
 Tr}\left[ {\cal M}\,(\Sigma+\Sigma^\dagger )\right]
\\
&=& 
\frac{1}{2}\partial_\mu \pi^0\, \partial^\mu \pi^0 +
\frac{1}{2}m_\pi^2(\pi^{0})^2  + \text{ self-interactions}
\\
&=& 
\frac{1}{2}\pi^0 
\left( -\partial^2 -m_\pi^2 \right)\pi^0  + \cdots,
\end{eqnarray}
where $\Sigma=\exp(2i\pi/f_\pi)$, ${\cal M}={\rm diag}(m_u,m_d)$
is the quark-mass matrix and $B_0$ is a constant.
The pion self-interaction terms are not needed at the order we are working so 
they are dropped, and we add the pion kinetic term in the last line above 
to Eq.~(\ref{append-L4}) to obtain our Lagrangian. 

However, this Lagrangian still includes the large 
scales $m_\pi$ and $\Delta_{}$, which must be integrated out of the theory. 
Since we are interested in a non-relativistic theory of pions we are 
justified in splitting the pion fields into creation and annihilation 
operators $\pi^0 = \hat \pi + \hat \pi^{\dagger}$. 
In addition we rescale the meson fields to make the large
scales explicit:
\begin{eqnarray}
 \hat \pi = \frac{1}{\sqrt{2m_\pi}} \, e^{-im_\pi t}\,  \pi \,,
\hspace{1cm}
 \hat \pi^\dagger = \frac{1}{\sqrt{2m_\pi}} \, e^{im_\pi t}\,\pi^\dagger \,,
\hspace{1cm}
{\bf D} \to e^{-im_\pi t}\,{\bf D}.
\label{pihat}
\end{eqnarray}
The pion kinetic term can then be expanded
\begin{eqnarray}
 {\cal L}_\pi 
&=& \frac{1}{2}\pi^0 
\left( -\partial_0^2 + \vec{\nabla}^2 -m_\pi^2 \right)\pi^0
\nn \\
&=&
{1\over 4 m_\pi} \left\{
 \pi^\dagger  \left( 2im_\pi\partial_0 - \partial_0^2 
+ \vec{\nabla}^2  \right)  \pi
+ \pi  \left( -2im_\pi\partial_0 - \partial_0^2 
+ \vec{\nabla}^2  \right)  \pi^\dagger 
\right. \nn \\
&& \left.
+ e^{-2im_\pi t} \pi \left( 2im_\pi\partial_0 - \partial_0^2 
+ \vec{\nabla}^2  \right)  \pi
+ e^{2im_\pi t} \pi^\dagger \left( -2im_\pi\partial_0 - \partial_0^2 
+ \vec{\nabla}^2  \right)  \pi^\dagger
\right\}
\nn \\
&=& 
\pi^\dagger \left( i\partial_0 + {\vec{\nabla}^2 \over 2m_\pi}
 \right) \pi + \text{higher-order relativistic corrections} \,.
 \end{eqnarray}
The terms in the third line include a large phase factor and as a consequence 
can be integrated out. 
In addition to modifying the pion propagator the field redefinition 
in Eq.~(\ref{pihat}) modifies the kinetic term for ${\bf D}$,
\begin{eqnarray}
 {\bf D}^\dagger \left(i\partial_0  
+ \frac{\vec{\nabla}^2}{2m_{D^*}}- \Delta_{} \right) {\bf D}
\to 
{\bf D}^\dagger \left(i\partial_0 
+ \frac{\vec{\nabla}^2}{2m_{D^*}}- \delta \right) {\bf D},
\end{eqnarray} 
where $\delta = \Delta_{} -m_\pi \simeq 7$ MeV.
Note that after the field rescaling the last term in Eq.~(\ref{append-L4}) 
contains a 
phase factor $e^{-im_\pi t}$, and can be dropped.

Finally, we obtain the the first three lines of the Lagrangian given 
in Eq.~(\ref{lag}) by 
another rephasing of the ${\bf D}$ and $\pi$ fields,
\begin{eqnarray}
 {\bf D} \to e^{-i \delta \, t}\, {\bf D}, \hspace{1cm} 
 \pi \to e^{-i \delta \, t} \, \pi.
\end{eqnarray}
This just shifts the residual mass from the ${\bf D}$ kinetic term
to the $\pi$ kinetic term. This last step is not essential, however it 
is convenient. The remaining terms in Eq.~(\ref{lag}) are short-distance 
interactions allowed by power counting and the symmetries of the theory.

\section{NLO diagrams}\label{nlo}

The NLO decay diagrams involving pion exchange are shown in 
Fig.~\ref{nlopiondecay}. 
The result for Fig.~\ref{nlopiondecay}a) is
\bea
i \frac{g^3 M^2_{DD^*}}{8\pi f^3_\pi(p_D^2 +\gamma^2)} \,
\bigg[ \left(\frac{2}{3}\Lambda_{\rm PDS} -h_1(p_D)\right) \,\vec{p}_\pi \cdot \vec{\epsilon}_X
+h_2(p_D)\,\vec{p}_D \cdot \vec{\epsilon}_X\,  \vec{p}_D \cdot \vec{p}_\pi \bigg] + (p_D \to p_{\bar D}) \, ,
\eea
where the functions $h_1(p)$ and $h_2(p)$ are given by:
\begin{align}
 h_1(p) &= \int_0^1 dx\sqrt{-p^2 x^2 +(p^2+\gamma^2 + \mu^2) x
 -\mu^2-i\epsilon} \nonumber \\
&= {p^2-\gamma^2-\mu^2 \over 4 p^2 } \gamma 
+ {(p^2+\gamma^2+\mu^2)^2-4p^2\mu^2 \over 8p^3} 
\tan^{-1}{2p\gamma\over - p^2+\gamma^2+\mu^2} \nonumber \\ 
&\quad -i{\mu\over 4p^2}(p^2+\gamma^2+\mu^2)-i 
 {(p^2+\gamma^2+\mu^2)^2-4p^2\mu^2 \over 16p^3} 
\ln{\gamma^2+(\mu-p)^2\over \gamma^2+(\mu+p)^2}
\\
 h_2(p) & =  \int_0^1 dx{x^2\over 
\sqrt{-p^2 x^2 +(p^2+\gamma^2 + \mu^2) x
 -\mu^2-i\epsilon} } \nonumber \\
&= -{5p^2+3\gamma^2+3\mu^2 \over 4 p^4 } \gamma 
+ {3(p^2+\gamma^2+\mu^2)^2-4p^2\mu^2 \over 8p^5} 
\tan^{-1}{2p\gamma\over - p^2+\gamma^2+\mu^2} \nonumber \\ 
&\quad -i{3 \mu\over 4p^4}(p^2+\gamma^2+\mu^2)-i 
 {3(p^2+\gamma^2+\mu^2)^2-4p^2\mu^2 \over 16p^5} 
\ln{\gamma^2+(\mu-p)^2\over \gamma^2+(\mu+p)^2} \, ,
\end{align}
and the result for Fig.~\ref{nlopiondecay}b) is
\bea
- \frac{g^3 M^2_{DD^*}}{12\pi f^3_\pi}\frac{\mu^3}{(p_D^2 +\gamma^2)^2} 
\, \vec{p}_\pi \cdot \vec{\epsilon}_X
 + (p_D \to p_{\bar D}).
\eea
Note that Fig.~\ref{nlopiondecay}b) includes, as a subgraph, the one-loop 
$D^*$ self-energy contribution. In the PDS scheme the self-energy graph 
has a linear divergence which gives an additive renormalization to the 
residual mass term of
the $D^{*}$. At tree level, we performed a field redefinition
which moved the residual mass term from  the kinetic term of the $D^*$ 
to the kinetic term of the pion
through a field redefinition. In order that loop corrections do not 
reintroduce a residual mass for the $D^{*}$, we introduce a  counterterm 
$-\delta_{\rm ct} (\bd^\dagger \bd +\bdbar^\dagger \bdbar)$, 
which is defined to cancel the residual mass term at each order in 
perturbation theory. At one-loop order, 
$\delta_{\rm ct} = g^2\mu^2\Lambda_{\rm PDS}/24\pi f_\pi^2$.
This linearly  divergent contribution to the self-energy also appears in
Fig.~\ref{sigmaNLO}b) and is canceled by an insertion of the residual
mass counterterm in a $D D^*$ bubble (not shown in the figure).

The NLO diagrams with the counterterms $C_2$ and $B_1$ are shown 
in Fig.~\ref{nlocontactdecay}. 
The result for Fig.~\ref{nlocontactdecay}a) is
\bea
- i C_2(\Lambda_{\rm PDS})\frac{g M_{DD^*}^2 
(\Lambda_{\rm PDS}-\gamma)}{4\pi f_\pi}
\, \frac{p_D^2-\gamma^2}{p_D^2+\gamma^2} \, \vec{p}_\pi \cdot 
\vec{\epsilon}_X + (p_D \to p_{\bar D}) \, ,
\eea
and the result for Fig.~\ref{nlocontactdecay}b) is
\bea
- i B_1(\Lambda_{\rm PDS})\frac{M_{DD^*}(\Lambda_{\rm PDS}-\gamma)}{2 \pi}
 \, \vec{p}_\pi \cdot \vec{\epsilon}_X \, .
\eea

Finally, we show the NLO wave function renormalization diagrams in 
Fig.~\ref{sigmaNLO}.
The NLO contribution to ${\rm Re} \,\Sigma^\prime(-E_X)$ from the graphs
in Fig.~\ref{sigmaNLO} is
\bea
\frac{g^2 M_{DD^*}^3}{12 \pi^2 f_\pi^2}
\bigg[ \frac{\Lambda_{\rm PDS}-\gamma}{\gamma} 
+\frac{2 \mu^2}{4\gamma^2+\mu^2} \bigg]
- C_2(\Lambda_{\rm PDS}) \frac{M_{DD^*}^3 (\gamma-\Lambda_{\rm PDS})
(2\gamma-\Lambda_{\rm PDS})}{2 \pi^2} \, .
\eea


\end{document}